\documentclass[twocolumn,aps,pra,showpacs,superscriptaddress]{revtex4}

\usepackage{graphicx}
\usepackage[intlimits]{amsmath}
\usepackage{amsfonts}
\usepackage{amssymb}
\usepackage{amsmath}
\usepackage{amscd}
\usepackage{latexsym}
\usepackage{bbm}
\usepackage[sort&compress]{natbib}
\usepackage{ifthen}
\usepackage{epsfig}
\usepackage{dcolumn}      
\usepackage{bm}           
\usepackage{amsfonts}
\usepackage{enumerate}
\usepackage[latin1]{inputenc} 

\newcommand{\bra}[1]{\ensuremath{\langle#1|}}

\newcommand{\ket}[1]{\ensuremath{|#1\rangle}}

\newcommand{\ketbra}[1]{\ensuremath{| #1 \rangle \langle #1 |}}

\newcommand{\eins}{\ensuremath{\openone}}
\newcommand{\HH}{\ensuremath{\mathcal{H}}}
\newcommand{\BB}{\ensuremath{\mathcal{B}}}
\newcommand{\WW}{\ensuremath{\mathcal{W}}}

\newcommand{\FF}{\ensuremath{\mathcal{F}}}

\newcommand{\BE}{\begin{equation}}
\newcommand{\EE}{\end{equation}}
\newcommand{\be}{\begin{equation}}
\newcommand{\ee}{\end{equation}}
\newcommand{\bea}{\begin{eqnarray}}
\newcommand{\eea}{\end{eqnarray}}
\newcommand{\kommentar}[1]{}

\newcommand{\mean}[1]{\ensuremath{\langle #1 \rangle}}

\renewcommand{\vr}{\ensuremath{\varrho}}
\newcommand{\II}{\ensuremath{\mathbbm I}}

\begin{document}
\title{Entanglement criteria based on local uncertainty
relations are strictly stronger than the computable cross norm
criterion}


\begin{abstract}

We show that any state which violates the computable cross norm (or
realignment) criterion for separability also violates the
separability criterion of the local uncertainty relations. The
converse is not true. The local uncertainty relations provide a 
straightforward construction of nonlinear entanglement witnesses 
for the cross norm criterion.

\end{abstract}


\author{Otfried G\"uhne}
\affiliation{Institut f\"ur Quantenoptik und Quanteninformation,
\"Osterreichische Akademie der Wissenschaften, A-6020 Innsbruck,
Austria}

\author{M\'aty\'as Mechler}
\affiliation{Research Group for Nonlinear and Quantum Optics,
Hungarian Academy of Sciences, University of P\'ecs, Ifj\'us\'ag
\'ut 6., H-7624 P\'ecs, Hungary}

\author{G\'eza T\'oth}
\affiliation{Research Institute for Solid State Physics and Optics,
Hungarian Academy of Sciences, P.O. Box 49, H-1525 Budapest,
Hungary} \affiliation{Max-Planck-Institut f\"ur Quantenoptik,
Hans-Kopfermann-Stra{\ss}e 1,
 D-85748 Garching, Germany}

\author{Peter Adam}
\affiliation{Research Group for Nonlinear and Quantum
Optics, Hungarian Academy of Sciences, University of P\'ecs,
Ifj\'us\'ag \'ut 6., H-7624 P\'ecs, Hungary}\affiliation{Research
Institute for Solid State Physics and Optics, Hungarian Academy of
Sciences, P.O. Box 49, H-1525 Budapest, Hungary}

\pacs{
03.67.-a,
03.65.Ud
}
\maketitle


Entanglement plays a central role in quantum information processing. 
Thus its characterization is important for the field: It is crucial 
to be able to decide whether or not a given quantum state is entangled. 
However, this so-called separability problem remains one of the most 
challenging unsolved problems in quantum physics.

Several sufficient conditions for entanglement are known. The first
of such criteria was the criterion of the positivity of the partial
transpose (PPT) \cite{ppt1}. This criterion is necessary and sufficient 
for $2\times 2$ and $2\times 3$ systems \cite{ppt2}, but in higher 
dimensional systems some entangled states escape the detection. The 
characterization of these PPT entangled states is thus of great interest. 
Recently, the computable cross norm (CCN) or realignment criterion was 
put forward by O.~Rudolph \cite{R02} and Chen and Wu \cite{CW03}. The 
original condition has been reformulated in several ways and extended to
multipartite systems \cite{R03,CW04,ccn1}. The CCN criterion allows
to detect the entanglement of many states where the PPT criterion
fails, however, some states which are detected by the PPT criterion,
cannot be detected by the CCN criterion \cite{R03}. In this way, one
may view the CCN criterion as complementary to the PPT criterion.
In addition to the  CCN criterion, there are also algorithmic 
approaches to the separability problem  which allow the detection of 
entanglement when the PPT criterion fails \cite{algos}.

A different approach to the separability problem tries to formulate 
separability criteria directly in mean values or variances of 
observables.
Typically, these conditions are 
formulated as Bell inequalities \cite{bell}, entanglement witnesses 
\cite{ppt2,entwit} or uncertainty relations 
\cite{nlin,lurs,H03,TG05,lurs2,ogprl}.
Here, the local uncertainty relations (LURs) by Hofmann and Takeuchi are 
remarkable \cite{lurs}. They have a clear physical interpretation and 
are quite versatile: It has been shown that they can be used to detect 
PPT entangled states \cite{H03}. It is further known that in certain 
situations they can provide a nonlinear refinement of linear entanglement 
witnesses \cite{TG05}. Consequently, the investigation of LURs has been 
undertaken in several directions \cite{lurs2, ogprl}.

In this paper we investigate the relation between the
CCN criterion and the LURs. We show that any state which
can be detected by the CCN criterion can also be detected
by a LUR. By providing counterexamples, we prove that the
converse does not hold. 
Our results show that the LURs can be viewed as nonlinear 
entanglement witnesses for the CCN criterion. In this way, 
we demonstrate a surprising  connection between  
permutation separability criteria (to which the CCN 
criterion belongs) \cite{ccn1}, criteria in terms of covariance 
matrices, such as LURs \cite{ogprl, cmcrit}, and the theory of 
nonlinear entanglement witnesses \cite{nlew, onlew}.
Further, in two Appendices we discuss the relation of our constructions
to other entanglement witnesses which have been proposed for
the CCN criterion and we calculate other nonlinear
entanglement witnesses for the CCN criterion \cite{nlew}.

Let us start by recalling the definition of separability.
A quantum state $\vr$ is called separable, if its density
matrix can be written as a convex combination of product
states,
\begin{equation}
\vr=\sum_k p_k \vr^{(A)}_k \otimes \vr^{(B)}_k,
\end{equation}
where $p_k \geq 0, \sum_k p_k = 1$ and $A$ and $B$ denote the two
subsystems. Throughout this paper, we denote by $\HH_A, \HH_B$ the
(finite dimensional) Hilbert spaces of Alice and Bob, and by 
$\BB(\HH_A), \BB(\HH_B)$ the
real vector space of the Hermitian observables on them. We first
assume that both $\HH_A$ and $\HH_B$ are $d$-dimensional, later we
discuss what happens if this is not the case.

The CCN criterion can be formulated in different ways. We use here a
formulation given in Ref.~\cite{R02} in Corollary 18, since it is best 
suited for our approach. It makes use of the Schmidt decomposition in 
operator space. Due to that, any density matrix $\vr$ can be written as
\begin{eqnarray}
\vr=\sum_k \lambda_k G^A_k \otimes G^B_k.
\label{rhodecompose}
\end{eqnarray}
where the $\lambda_k \geq 0$ and $G^A_k$ and $G^B_k$ are
orthogonal bases of the observable spaces $\BB(\HH_A)$
resp. $\BB(\HH_B).$ Such a basis consists of $d^2$
observables which have to fulfill
\begin{eqnarray}
Tr(G^A_k G^A_l) = Tr(G^B_k G^B_l)=\delta_{kl}.
\label{localorthogonal}
\end{eqnarray}
We refer to such observables as {\it local orthogonal
observables} (LOOs) \cite{YL05}. For instance, for qubits
the (appropriately normalized) Pauli matrices together with
the identity form a set of LOOs (see Eq. (\ref{AkBk})). Note that, given a set
$G^A_k$ of LOOs, any other set $\tilde{G}^A_l$ of LOOs
is of the form $\tilde{G}^A_l = \sum_k O_{lk} G^A_k,$
where  $O_{lk}$ is a $d^2 \times d^2$ real orthogonal
matrix \cite{YL05}.

As for the usual Schmidt decomposition, the $\lambda_k$
are (up to a permutation) unique and if the  $\lambda_k$
are pairwise different, the $G^A_k$ and $G^B_k$ are also
unique (up to a sign). The $\lambda_k$ can be computed
as in the Schmidt decomposition: First, one decomposes
$\vr=\sum_{kl} \mu_{kl} \tilde{G}^A_k \otimes \tilde{G}^B_l$
with arbitrary LOOs $\tilde{G}^A_k$ and $\tilde{G}^B_l,$
then, by performing the singular value decomposition of
$\mu_{kl}$ one arrives at Eq.~(\ref{rhodecompose}), the
$\lambda_k$ are the roots of the eigenvalues of the matrix
$\mu \mu^\dagger.$

The CCN criterion states that if $\vr$ is separable, then the sum
of all $\lambda_k$ is smaller than one:
\be
\mbox{ $\vr$ is separable } \Rightarrow \sum_k \lambda_k \leq 1.
\ee
Hence, if $\sum_k \lambda_k > 1$ the state must be entangled.
For states violating this criterion, an entanglement witness
can directly be written down. Recall that an entanglement
witness $\WW$ is an observable  with a positive expectation
value on all separable states, hence a negative expectation
value signals the presence of entanglement \cite{entwit}.
Given a state in the form (\ref{rhodecompose}) which violates
the CCN criterion, a witness is given by  \cite{remark1}
\be
\WW = \eins - \sum_k G^A_k \otimes G^B_k,
\label{ccnwit}
\ee
since for this state we have $Tr(\WW \vr) = 1- \sum_k
\lambda_k < 0$ due to the properties of the LOOs. On the other
hand, if  $\vr=\sum_{kl} \mu_{kl} {G}^A_k \otimes {G}^B_l$
were separable, then $Tr(\WW \vr ) = 1 - \sum_k \mu_{kk}
\geq 1 - \sum_k \lambda_k \geq 0,$ since $\sum_k \mu_{kk}
\leq \sum_k \lambda_k$ due to the properties of the singular
value decomposition \cite{hornjohnson}. It is clear that any
state violating the CCN criterion can be detected by a witness
of the type (\ref{ccnwit}). Note that other forms of entanglement
witnesses for the CCN criterion have also been proposed
\cite{CW04}, we will discuss them in the Appendix A.

Let us now discuss the LURs. This criterion is formulated as
follows: Given some non-commuting observables $A_k$ on Alice's space
and $B_k$ on Bob's space, one may compute  strictly positive numbers
$C_A$ and $C_B$ such that \be \sum_{k=1}^n \Delta^2 (A_k) \ge C_A,
\;\;\;\; \sum_{k=1}^n \Delta^2 (B_k) \ge C_B \ee holds for all
states for Alice, resp.~Bob. Here, $\Delta^2 (A)
=\mean{A^2}-\mean{A}^2$ denotes the variance of an observable $A.$
Then it can be proved that for separable states
\begin{eqnarray}
\sum_{k=1}^n \Delta^2 (A_k \otimes \openone + \openone \otimes B_k)
&\ge& C_A +C_B \label{entcond}
\end{eqnarray}
has to hold. Any quantum state which violates Eq.~(\ref{entcond})
is entangled. Physically, Eq.~(\ref{entcond}) may be interpreted
as stating that separable states always inherit the uncertainty
relations which hold for their reduced states \cite{eur}.

To connect the LURs with the CCN criterion, first note that for any
LOOs $G^A_k$ the relation \be \sum_{k=1}^{d^2} \Delta^2 (G^A_k) \ge
d-1, \label{lur1} \ee holds. This can be seen as follows. If we
choose the $d^2$ LOOs 
\be
G^A_k = 
\left\{
\begin{split}
&\frac{1}{\sqrt{2}}(\ket{m}\bra{n}+\ket{n}\bra{m}),
\\
&\;\; \mbox{for }1 \leq k \leq (d(d-1))/2; \;\;\; 1 \leq m < n \leq d;
\\
&\frac{1}{\sqrt{2}}(i \ket{m}\bra{n}-i \ket{n}\bra{m}),
\\
&\;\;\mbox{for }(d(d-1))/2 < k \leq  (d(d-1)); 
\\ &\;\;\mbox{and } 1 \leq m < n \leq d;
\\
&\ket{m}\bra{m}
\;\;\;\;\;
\mbox{for }{d(d-1)}< k \leq d^2; \;\;\; 1\leq m \leq d; \;\;\;
\end{split}
\right.
\nonumber
\ee
one can directly calculate that $\sum_k (G^A_k)^2 = d \openone$
and that $\sum_k \mean{G^A_k}^2 = Tr(\vr^2) \leq 1.$ For general
$\tilde{G}^A_k = \sum_l O_{kl} G^A_l$ we have
$\sum_k (\tilde{G}^A_k)^2 = \sum_{klm} O^T_{lk} O_{km} G^A_l G^A_m
= d \openone $ since $O$ is orthogonal and again
$\sum_k \mean{\tilde{G}^A_k}^2 = Tr(\vr^2) \leq 1$ \cite{rem2}.
Similarly, we have for Bob's system
\be
\sum_{k=1}^{d^2} \Delta^2 (-G^B_k) \ge  d-1,
\label{lur2}
\ee
where the minus sign has been inserted for later convenience.

Combining Eqs.~(\ref{lur1}, \ref{lur2}) with the method of the
LURs, using the fact that $\sum_k (G^A_k)^2 = \sum_k (G^B_k)^2
= d \openone$  one can directly calculate that for separable states
\be
1-\sum_k \mean {G^A_k \otimes G^B_k} -
\frac{1}{2}
\sum_k \mean {G^A_k \otimes \openone  - \openone \otimes G^B_k}^2
\geq 0.
\label{lur3}
\ee
The first, linear part is just the expectation value of the
witness (\ref{ccnwit}), from this some positive terms are
subtracted. Since  any state which violates the CCN criterion
can be detected by the witness in Eq.~(\ref{ccnwit}) it can
also be detected by the LUR in Eq.~(\ref{lur3}) and we have:

{\bf Theorem.} Any state which violates the computable cross norm
criterion can be detected by a local uncertainty relation, while the
converse is not true.

To prove the second statement of the theorem we will later give
explicit counterexamples of states which can be detected by
a LUR, but not by the CCN criterion. Before doing that, let us
add some remarks.

First, the Theorem from above can be interpreted in the following
way: While the witness in Eq.~(\ref{ccnwit}) is the natural linear
criterion for states violating the CCN criterion, the LUR in
Eq.~(\ref{lur3}) is the natural {\it nonlinear} witness for these
states. The fact that LURs can sometimes be viewed as nonlinear
witnesses which improve linear witnesses has been observed before
\cite{TG05}. The theorem, however, proves that the LURs provide
{\it in general} improvements for witnesses of the type
(\ref{ccnwit}). Note, that there are other possible nonlinear
improvements on these witnesses as discussed in Appendix B.

Second, we have to discuss what happens if the dimensions of the
Hilbert spaces $\HH_A$ and $\HH_B$ are not the same. So let us
assume that $d_A = dim(\HH_A) < d_B = dim(\HH_B).$ Then, in Eq.
(\ref{rhodecompose}) there are $d_A^2$ different $G^A_k$ and
$G^B_k.$ The $G^A_i$ form already a set of LOOs for $\HH_A$ and one
can find further $d_B^2 -d_A^2$ observables $G^B_k$ to complete the
set $\{G^B_k\}$ to become a complete set of LOOs for $\HH_B$. Using
then the LURs with the definition $G^A_k = 0$ for
$k=d_A^2+1,...,d_B^2$ proves the claim.

Now we present two examples which show that the LURs are strictly stronger 
than the CCN criterion. 
First, let us consider a noisy singlet state of the form
\begin{eqnarray}
\varrho_{\rm ns}(p):=p\ketbra{\psi_{\rm s}}+(1- p) \varrho_{\rm
sep}, \label{rhon2}
\end{eqnarray}
where the singlet is $ \ket{\psi_{\rm s}}:= (\ket{01}-\ket{10})/{\sqrt{2}},$
and the separable noise is given as
$\varrho_{\rm sep}:=2/3\ketbra{00}+1/3\ketbra{01}.$  Using the PPT criterion
one can see that the state is entangled for any $p>0.$ First we check
for which values of $p$ the state $\varrho_{\rm ns}$ is detected as
entangled by the CCN criterion. It can be seen that $\varrho_{\rm ns}(p)$ 
violates the CCN criterion for all $p> 0.292.$
Now we define $G^A_k$ and $G^B_k$ as
\begin{eqnarray}
\{G^A_k\}_{k=1}^4
&=&
\{
-\frac{\sigma_x}{\sqrt{2}},
-\frac{\sigma_y}{\sqrt{2}},
-\frac{\sigma_z}{\sqrt{2}},
\frac{\openone}{\sqrt{2}}
\}, \nonumber\\
\{G^B_k\}_{k=1}^4
&=&
\{\frac{\sigma_x}{\sqrt{2}},
\frac{\sigma_y}{\sqrt{2}},
\frac{\sigma_z}{\sqrt{2}},
\frac{\openone}{\sqrt{2}}
\}. \label{AkBk}
\end{eqnarray}
These $G^A_k$ and $G^B_k$ are the matrices corresponding to the
Schmidt decomposition of $\ketbra{\psi_{\rm s}}.$ Using
Eq.~(\ref{lur3}) with these LOOs one finds that $\varrho_{\rm ns}$
is detected as entangled by the LURs at least for $p> 0.25.$

For the second example, we consider the $3\times 3$ bound
entangled state defined in \cite{terhal} mixed with white noise:
\begin{align}
\ket{\psi_0} &= \frac{1}{\sqrt{2}}\ket{0}(\ket{0}-\ket{1}),
\;\;\;
\ket{\psi_1}=\frac{1}{\sqrt{2}}(\ket{0}-\ket{1})\ket{2},
\nonumber\\
\ket{\psi_2}&=\frac{1}{\sqrt{2}}\ket{2}(\ket{1}-\ket{2}),
\;\;\;
\ket{\psi_3}=\frac{1}{\sqrt{2}}(\ket{1}-\ket{2})\ket{0},
\nonumber\\
\ket{\psi_4}&=\frac{1}{3}(\ket{0}+\ket{1}+\ket{2})(\ket{0}+\ket{1}+\ket{2}),
\nonumber\\
\varrho_{BE}&=\frac{1}{4}(\eins-\sum_{i=0}^4\ket{\psi_i}\bra{\psi_i});
\;\;\;\; \vr(p) = p \varrho_{BE} + (1-p)\frac{\eins}{9}. \nonumber
\end{align}
The states $\vr(p)$ are detected as entangled via the CCN criterion
whenever $p > p_{\rm ccn} = 0.8897.$ Taking the LUR (\ref{lur3}) with
the Schmidt matrices of $\vr(p_{\rm ccn})$ as LOOs, one finds that
the states $\vr(p)$ must already be entangled for $p > p_{\rm lur} = 0.8885.$
Thus, the LURs are able to detect states which are neither detected by the CCN
criterion, nor by the PPT criterion. Note that $\vr(p)$ 
is known to be entangled  at least for $p> 0.8744$ \cite{CW04}.

In conclusion, we showed that entanglement criteria based on
local uncertainty relations are strictly stronger than the
CCN criterion. The local uncertainty relations can be
viewed as the natural nonlinear entanglement witnesses for
the CCN criterion. The question, whether there is also a relation
between the PPT criterion and local uncertainty relations is
very interesting. We leave this problem for future research.

We thank H.J. Briegel, M. Lewenstein, N. Lütkenhaus, 
M. Piani and M.M. Wolf for helpful
discussions. We acknowledge the support of the European Union (Grant
Nos. MEIF-CT-2003-500183 and MERG-CT-2005-029146, OLAQI, PROSECCO,
QUPRODIS, RESQ, SCALA), the FWF, the DFG, the Kompetenznetzwerk
Quanteninformationsverarbeitung der Bayerischen Staatsregierung and
the National Research Fund of Hungary  OTKA under contracts T049234
and T043287.

\appendix

\section{Connection to the witnesses proposed in Ref.~\cite{CW04}}

Now we show that the entanglement witness defined in
Eq.~(\ref{ccnwit}) is identical to the witness defined in
Ref.~\cite{CW04} based on a different formulation of the CCN
criterion. Let us first review the realignment map.
For a density matrix $\vr = \sum_{kl} \mu_{kl} {G}^A_k \otimes
{G}^B_l$ the realigned matrix is given by \cite{R02}
\begin{equation}
R(\vr):=\sum_{kl} \mu_{kl} \ket{G^A_k} \bra{G^B_l}
\label{Rrho}
\end{equation}
Here $\ket{G^A_k}$ denotes a column vector obtained from $G^A_k$ by
joining its columns consecutively while $\bra{G^B_k}$ denotes the
transposition of a column vector obtained similarly from $G^B_k.$
$R(\vr)$ can also be computed by a reordering (``realignment'') of
the matrix entries of $\vr,$ as explained in Ref.~\cite{CW03}. The
CCN criterion states that if $\|R(\vr)\|_1>1$ then $\rho$ is
entangled \cite{R02,R03,CW03,CW04}. Here $\|A\|_1$ denotes the trace
norm, i.e., the sum of the singular values of matrix $A.$ If $\vr
=\sum_k \lambda_k A_k \otimes B_k $ is given in its Schmidt
decomposition, we have $R(\vr)=\sum_{k} \lambda_{k} \ket{A_k}
\bra{B_k} $ and $\|R(\rho)\|_1=\sum_k \lambda_k.$
In this case  $R(\vr)$ is already given in its singular value
decomposition. To make this even more transparent,
let us define $\Sigma={\rm diag}(\lambda_1,\lambda_2,...)$,
$U=[\ket{A_1},\ket{A_2},...]$ and $V=[\ket{B_1},\ket{B_2},...].$
Then we obtain the decomposition $R(\vr)=U\Sigma V^\dagger.$

Now we can show that the witness Eq.~(\ref{ccnwit}) can be rewritten
using the inverse of $R.$ For that we need to observe that $\sum_k
A_k \otimes B_k =
R^{-1}(\sum_k\ket{A_k}\bra{B_k})=R^{-1}(UV^\dagger).$ Hence the
witness Eq.~(\ref{ccnwit}) can be written as
\begin{equation}
\WW=\eins - R^{-1}(UV^\dagger).
\label{app2}
\end{equation}
Since $R$ realigns the matrix entries, we have always $R^{-1}(X^*)=
R^{-1}(X)^*.$ Furthermore, since $\sum_k A_k \otimes B_k$ is
Hermitian, $R^{-1}(UV^\dagger)$ is also Hermitian. Thus the witness
in Eq.~(\ref{app2}) can be written as $\WW=\eins-[R^{-1}(U^*
V^T)]^T,$ which is the witness presented in Ref.~\cite{CW04}.

\section{More nonlinear witnesses}

Recently, a method to calculate nonlinear improvements
for a given general witness has been developed \cite{nlew}.
Here, we apply this method to Eq.~(\ref{ccnwit}). 

To start, we first have to calculate the positive map $\Lambda:
\BB(\HH_A) \rightarrow \BB(\HH_B)$ corresponding to $\WW$
\cite{jamiol}. This is 
$ 
\Lambda(\vr) = Tr_A [\WW(\vr^T\otimes \eins_B)],
$
and one can directly see that for
$\vr=\sum_i \alpha_i (G^A_i)^T$ we have $\Lambda(\vr)=
Tr(\vr)\eins_B -\sum_i \alpha_i (G^B_i).$ We can assume without the
loss of generality that $d\Lambda$ is trace non-increasing, otherwise 
we rescale $\WW$ to obtain this. According to the 
Jamio{\l}kowski isomorphism the witness can then be rewritten as 
\be 
\WW = (\II_{A} \otimes d \Lambda)(\ketbra{\phi^+}), 
\label{nl15} 
\ee 
where $\ket{\phi^+}=
\sum_i \ket{ii}/\sqrt{d}$ is a maximally entangled state on $\HH_{A}
\otimes \HH_A.$ Since for LOOs $\sum_i Tr(G^A_i)G^A_i = \eins$
holds, Eq.~(\ref{nl15}) implies that $\ketbra{\phi^+}=\sum_i G^A_i
\otimes (G^A_i)^T/d.$

To write down a nonlinear improvement, we can take an arbitrary
state $\ket{\psi} \in \HH_{A} \otimes \HH_A$ which has a maximal
squared Schmidt coefficient $s(\psi).$ Then, defining $X=(\II_{A}
\otimes d \Lambda)(\ket{\phi^+}\bra{\psi})$ the functional 
\be
\FF(\vr) = \mean{\WW} - \mean{X} \mean{X^\dagger}/{s(\psi)}
\ee 
is a nonlinear improvement of $\WW$ \cite{nlew}.

To give a first example, let us choose an arbitrary unitary $U^A$ on
$\HH_A$ and define $\ket{\psi}= (U^A)^\dagger \otimes \eins
\ket{\phi^+},$ which implies that $s(\psi) =1/d.$ Then direct
calculations lead to the nonlinear witness \be \FF(\vr) = \mean{\WW}
- d  \mean{\WW (U^A \otimes \eins)} \mean{(U^A \otimes
\eins)^\dagger \WW}. \ee

To give a second example, let us define $\ket{\psi}=\eins \otimes
(U^A)^\dagger \ket{\phi^+}.$ Using the coefficients
$\eta_{ij}=Tr[(G^A_i)^T (G^A_j)^T U^A]$ we can directly calculate
that $X = (\II_{A} \otimes \Lambda) (\sum_i G^A_i \otimes (G^A_i)^T
U^A) = \eins - \sum_{ij}  G^A_i \otimes \eta_{ij} G^B_j. $ Hence,
\be \FF(\vr) = \mean{\WW} - d  \mean{\eins - \sum_{ij} G^A_i \otimes
\eta_{ij} G^B_j} \mean{\eins - \sum_{ij}  G^A_i \otimes \eta_{ij}^*
G^B_j} \nonumber \ee is another nonlinear witness, improving the
witness in Eq.~(\ref{ccnwit}). The structure of these witnesses is
quite different from the structure of the LURs. 
Thus other nonlinear
witnesses can be  derived for the CCN criterion, which do not
coincide with the LURs.



\begin{thebibliography}{99}
\bibitem{remark1} This witness has also been derived in a different
formulation in Ref.~\cite{YL05}.
\end{thebibliography}
\end{document}